\documentclass[showpacs,twocolumn,amsmath,amssymb,pre,aps,dvips]{revtex4} %% 4-pages !!
\usepackage{graphicx,epsfig} % Include figure files
\usepackage{bm} % bold math

\begin{document}

\title{Transition from single-file to two-dimensional diffusion of interacting particles in a quasi-one-dimensional channel}

\author{
D.~Lucena$^{1,2}$, D.~V.~Tkachenko$^{2}$, K.~Nelissen$^{1,2}$, V.~R.~Misko$^{2}$,
W.~P.~Ferreira$^{1}$, G.~A.~Farias$^{1}$, and F.~M.~Peeters$^{1,2}$
}

\affiliation{
$^{1}$Departamento de F\'isica, Universidade Federal do Cear\'a, Caixa Postal 6030,
Campus do Pici, 60455-760 Fortaleza, Cear\'a, Brazil \\
$^{2}$Department of Physics, University of Antwerp, Groenenborgerlaan 171,
B-2020 Antwerpen, Belgium
}

\date{\today}

\begin{abstract}
Diffusive properties of a monodisperse system of interacting particles confined
to a \textit{quasi}-one-dimensional (Q1D) channel are studied using
molecular dynamics (MD) simulations.
We calculate numerically the mean-squared displacement (MSD) and investigate the
influence of the width of the channel (or the strength of the confinement potential)
on diffusion in finite-size channels of different shapes (i.e., straight and circular).
The transition from single-file diffusion (SFD)
to the two-dimensional diffusion regime is investigated.
This transition (regarding the calculation of the scaling exponent ($\alpha$) of the MSD $\langle \Delta x^{2}(t) \rangle$ $\propto t^{\alpha}$) as a function of the width of the channel, is shown to change depending on the channel's confinement profile. In particular the transition can be either smooth (i.e., for a parabolic confinement potential) or rather sharp/stepwise (i.e., for a hard-wall potential), as distinct from infinite channels where this transition is abrupt.
This result can be explained by qualitatively different distributions
of the particle density for the different confinement potentials.
\end{abstract}

\pacs{05.40.-a, 66.10.C-, 82.70.Dd, 83.10.Rs}

\maketitle

\section{Introduction}

There is a considerable theoretical and practical interest in the dynamics of systems
of interacting particles in confined geometries \cite{bocquet}.
Single-file diffusion (SFD) refers to a one-dimensional (1D) process where the motion
of particles
in a narrow channel (e.g., \textit{quasi}-1D systems) is limited such that particles
are not able to cross each other.
As a consequence, the system diffuses as a whole resulting in anomalous diffusion.
The mechanism of SFD was first proposed by Hodgkin and Keynes \cite{hodgkin} in order
to study the passage of molecules through narrow pores.
Since the order of the particles is conserved over time, this results in unusual
dynamics of the system \cite{kargerPRA,martin}, different from what is predicted from diffusion governed by Fick's law.
The main characteristic of the SFD phenomena is that, in the long-time limit, the MSD (mean-square displacement, defined as $\langle \Delta x^{2}(t) \rangle = \langle \sum_{i=1}^{N} (1/N) [x_{i}(t+\Delta t) - x_{i}(t)]^{2} \rangle_{\Delta t}$) scales with time as
\begin{equation}
\langle \Delta x^{2}(t) \rangle \propto t^{0.5}.
\end{equation}
This relation was first obtained analytically in the pioneering work of Harris \cite{TEHarris}.
Recent advances in nanotechnology have stimulated a growing interest in SFD,
in particular, in the study of transport in nanopores \cite{soaresPRB,soaresPRL}.
Ion channels of biological membranes and carbon nanotubes \cite{GHummer} are examples
of such nanopores.
The macroscopic flux of particles through such nanopores is of great importance for
many practical applications, e.g., particle transport across membranes is a crucial
intermediate step in almost all biological and chemical engineering processes.
SFD was observed in experiments on diffusion of molecules in zeolite molecular
sieves \cite{meier}.
Zeolites with unconnected parallel channels may serve as a good realization of the
theoretically investigated one-dimensional systems. SFD is also related to growth
phenomena \cite{grphen}.

The theoretical background of SFD was developed in early studies on transport phenomena
in 1D channels \cite{lebowitz,levitt,richards}. It is also interesting to learn how
the size of the system will influence the diffusive properties of the system.
SFD in finite size systems has been the focus of increasing attention since there
are few exact theoretical results to date \cite{PRE80-051103,PRE80-031118,PRL102-050602},
which showed the existence of different regimes of diffusion.

Colloidal systems, complex plasmas and vortex matter in type-II superconductors are
examples
of systems where SFD may occur.
The use of colloidal particles is technically interesting since
it allows real time and
spatial direct observation of their position, which is a great advantage as compared
to atoms or molecules, as shown recently in, e.g., the experimental study of defect
induced melting \cite{alsayed309}. One typically uses micro-meter size colloidal
particles in narrow channels, as shown in \cite{weiScience,PRLsfd}.
The paramagnetic colloidal spheres of 3.6 $\mu m$ were confined in
circular trenches fabricated by photolithography and their trajectories were followed
over long periods of time.
Several other studies have focused on the diffusive properties of complex plasmas.
A complex plasma consists of micrometer-sized (``dust'') particles immersed in a
gaseous plasma
background.
Dust particles typically acquire a negative charge of several thousand elementary
charges,
and thus they interact with each other through their strong electrostatic
repulsion \cite{gioPRE}.

Systems of particles moving in space of reduced dimensionality or submitted to an
external confinement potential exhibit different behavior from their free-of-border
counterparts \cite{wandPRB}.
The combined effect of interaction between the particles and the confinement potential
plays a crucial role in their physical and chemical properties \cite{kwinPRE}.
In Ref.~\cite{kwinEPL80}, it was found that SFD depends on the
inter-particle interaction and can even be suppressed if the
interaction is sufficiently strong, resulting in a slower subdiffusive behavior,
where $\langle \Delta x^{2}(t) \rangle$ $\propto t^{\alpha}$, with $\alpha < 0.5$.

In this paper, we will investigate the effects of confinement potential on the
diffusive properties of a Q1D system of interacting particles.
In the limiting case of very narrow (wide) channels, particle diffusion can be referred
to SFD (2D regime) characterized by a subdiffusive (normal diffusive) long-time regime
where the mean-squared displacement (MSD) $\langle \Delta x^{2}(t) \rangle$ $\propto t^{0.5}$ ($\propto t^{1.0}$).
Recall that the MSD of a tagged hard-sphere particle in a one
dimensional infinite system is characterized by two limiting diffusion behaviors: for time scales
shorter than a certain crossover time $\tau_{c} = 1/D\rho^{2}$, where $D$ is the
diffusion coefficient and $\rho$ is the particle concentration, $\langle \Delta x^{2}(t) \rangle$ 
$\propto t^{1.0}$ which is referred to as the normal diffusion regime \cite{underdamped}.
For times larger than $\tau_{c}$, the system exhibits a subdiffusive behavior,
with the MSD $\langle \Delta x^{2}(t) \rangle$ $\propto t^{0.5}$, which characterizes the single-file diffusion regime.
Between these two regimes, there is a transient regime exhibiting a non-trivial
functional form.

However, in case of a \textit{finite} system of diffusing particles (e.g., a circular chain
or a straight chain in the presence of periodic boundary conditions), the SFD regime (i.e., with $\langle \Delta x^{2}(t) \rangle$ $\propto t^{0.5}$)
does not hold for $t \rightarrow \infty$, unlike in an infinite system. Instead, for sufficiently long times, the
SFD regime turns to the regime of \textit{collective} diffusion, i.e., when the whole system diffuses as a single ``particle'' with
a renormalized mass. This diffusive behavior has been revealed in experiments \cite{weiScience,MSJ} and theoretical studies \cite{Tepper,kwinEPL80,submitEPL,SJNew,PMCentres}.
This collective diffusion regime is similar to the initial short-time diffusion regime and it is characterized by either $\langle \Delta x^{2}(t) \rangle$ $\propto t^{1.0}$,
for overdamped particles (see, e.g., \cite{Tepper,weiScience,kwinEPL80}) or by $\langle \Delta x^{2}(t) \rangle$ $\propto t^{2.0}$ (followed by the MSD $\propto t^{1.0}$), for underdamped systems \cite{submitEPL,SJNew}. Correspondingly, the time interval where the SFD regime is observed becomes \textit{finite} in finite size systems. It depends on the lenght of the chain of diffusing particles: the longer the chain the longer the SFD time interval. Therefore, in order to observe a clear power-law behavior (i.e., $\langle \Delta x^{2}(t) \rangle$ $\propto t^{\alpha}$) one should consider sufficiently large systems.

Here we focus on this intermediate diffusion regime and we show that it can be characterized by $\langle \Delta x^{2}(t) \rangle$ $\propto t^{\alpha}$, where $0.5 < \alpha < 1.0$, depending on the width (or the strenght of the confinement potential) of the channel.
We analyze the MSD for two different channel geometries:
(i) a linear channel,
and
(ii) a circular channel.
These two systems correspond to different experimental realizations of diffusion
of charged particles in narrow channels \cite{weiScience,rndWalkToSFD}.
The latter one (i.e., a circular channel) has obvious advantages:
(i) it allows a long-time observation of diffusion using a
relatively short circuit, and
(ii) it provides constant average particle density and absence
of density gradients (which occur in, e.g., a linear
channel due to the entry/exit of particles in/from the channel).
Thus circular narrow channels were used in diffusion experiments
with colloids \cite{weiScience} and metallic charged particles (balls) \cite{MSJ}.
Furthermore, using different systems allows us to demonstrate that
the results obtained in our study are generic and do not depend on
the specific experimental set-up.

This paper is organized as follows.
In Sec.~II, we introduce the model and numerical approach.
In Sec.~III diffusion in a system of interacting particles,
confined to a straight hard-wall or parabolic channel,
is studied as a function of the channel width or confinement strength.
In Sec.~IV, we discuss the possibility of experimental observation
of the studied crossover from the SFD to 2D diffusive regime.
For that purpose, we analyze diffusion in a realistic experimental
set-up, i.e., diffusion of massive metallic balls embedded in
a circular channel with parabolic confinement whose strength can
be controlled by an applied electric field. 
The long-time limit is analyzed in Sec.~V using a discrete site model. 
Finally, the conclusions are presented in Sec.~VI.

\section{Model system and numerical approach}

Our model system consists of $N$ identical charged particles interacting through a repulsive pair potential $V_{int}(\vec{r}_{ij})$.
In this study, we use a screened Coulomb potential (Yukawa potential),  $V_{int} \propto \exp(-r/\lambda_{D})/r$.
In the transverse direction, the motion of the particles is restricted either by a hard-wall or by a parabolic confinement potential.
Thus the total potential energy of the system can be written as:
\begin{equation}\label{eq1}
H = \sum_{i=1}^{N} V_{c}(\vec{r}_{i}) + \sum_{i>j=1}^{N} V_{int}(\vec{r}_{ij}).
\end{equation}
The first term in the right-hand side (r.h.s.) of Eq.~(\ref{eq1}) represents the confinement potential, where $V_{c}(\vec{r}_{i})$ is given by:
\begin{equation}\label{eq2}
V_{c}(\vec{r}_{i}) = \left\{
\begin{array}{ll} 0 & \mbox{for } |y_{i}|\leq R_{w}/2  \\
\infty & \mbox{for } |y_{i}| > R_{w}/2,
\end{array}
\right.
\end{equation}
for the hard-wall confinement,
\begin{equation}\label{vc1d}
V_{c}(\vec{r}_{i}) = \frac{1}{2}m\omega^ {2}_{0} y^{2}_{i},
\end{equation}
for parabolic one-dimensional potential (in the $y$-direction), and by
\begin{equation}
V_{c}(\vec{r}_{i}) = \beta(r_{0} - r_{i})^{2},
\label{vc}
\end{equation}
for parabolic circular confinement.
Here $R_{w}$ is the width of the channel (for the hard-wall potential), $m$ is the mass of the particles, $\omega_{0}$ is the strenght of the parabolic 1D confining potential, $r_{0}$ is the coordinate of the minimum of the potential energy
and $r_{i}$ is the displacement of the $i$th particle from $r_{0}$
(for the parabolic circular potential).
Note that in case of a circular channel,
$r_{0}=r_{ch}$, where
$r_{ch}$ is the radius of the channel.

The second term in the r.h.s. of Eq. (\ref{eq1}) represents the interaction potential between the particles.
For the screened Couloumb potential,
\begin{equation}\label{eq3}
V_{int}(\vec{r}_{ij}) = \frac{q^{2}}{\epsilon} \frac{e^{- |\vec{r}_{i}-\vec{r}_{j}| / \lambda_{D} }}{|\vec{r}_{i}-\vec{r}_{j}|},
\end{equation}
where $q$ is the charge of each particle,
$\epsilon$ is the dieletric constant of the medium,
$r_{ij} = |\vec{r}_{i} - \vec{r}_{j}|$ is the distance between
$i$th and $j$th particles, and $\lambda_{D}$ is the Debye screening length.
Substituting (\ref{eq3}) into Eq. (\ref{eq1}), we obtain the potential energy of the system $H_{Y}$:
\begin{equation}\label{eq4}
H_{Y} = \sum_{i=1}^{N} V_{c}(\vec{r}_{i}) + \frac{q^{2}}{\epsilon} \sum_{i>j=1}^{N} \frac{e^{- |\vec{r}_{i} - \vec{r}_{j}|/\lambda_{D}}}{|\vec{r}_{i} - \vec{r}_{j}|}.
\end{equation}
In order to reveal important parameters which characterize the system, we rewrite the energy $H_{Y}$ in a dimensionless ($H'_{Y}$) form
by making use of the following variable transformations:
$H_{Y} = (q^{2}/\epsilon a_{0})H'_{Y}$, $r = r'a_{0}$, where $a_{0}$ is the mean inter-particle distance.
The energy of the system then becomes
\begin{equation}\label{eq5}
H'_{Y} = \sum_{i=1}^{N} V'_{c}(\vec{r'}_{i}) + \sum_{i>j=1}^{N} \frac{e^{-\kappa |\vec{r'}_{i} - \vec{r'}_{j}|}}{|\vec{r'}_{i}
- \vec{r'}_{j}|},
\end{equation}
where $\kappa = a_{0}/\lambda_{D}$ is the screening parameter of the interaction potential. In our simulations in Sec. III, we use a typical value of $\kappa = 1.0$ for colloidal systems
and $\lambda_{D} = 10^{-5}$m.

The hard-wall confinement potential is written as
\begin{equation}\label{eq6}
V'_{c}(\vec{r'}_{i}) = \left\{
\begin{array}{ll} 0 & \mbox{for } |y'_{i}|\leq R^{'}_{w}/2  \\
\infty & \mbox{for } |y'_{i}| > R^{'}_{w}/2,
\end{array}
\right.
\end{equation}
where $R^{'}_{w}$ is scaled by the inter-particle distance $a_{0}$. We also introduce a dimensionless parameter
\begin{equation}\label{ki}
\chi = \frac {m(\omega_{0}a_{0})^{2}}{2k_{B}T},
\end{equation}
which is a measure of the strenght of the parabolic 1D confinement potential.

For colloidal particles moving in a nonmagnetic liquid, their motion
is overdamped and thus the stochastic Langevin equations of motion can be
reduced to those for Brownian particles \cite{ermak}:
\begin{eqnarray}\label{eq7}
\frac{d\vec{r}_{i}}{dt} = \frac{D_{i}}{k_{B}T} \Big[ - \sum_{j \neq i} \vec{\nabla}_{i} V_{int}(\vec{r}_{ij}) 
\nonumber \\ 
- \vec{\nabla}_{i} V_{c}(\vec{r}_{i}) + \vec{F}^{i}_{T}(t) \Big].
\end{eqnarray}
Note, however, that in Sec.~IV we will deal with massive metallic
balls and therefore we will keep the inertial term in the Langevin
equations of motion.

In Eq. (\ref{eq7}), $\vec{r}_{i}$,
$D_{i}$ and $m_{i}$ are the position, the self-diffusion
coefficient (measured in m$^{2}$/s) and the mass (in kg) of the $i$th particle, respectively, $t$ is the time (in seconds), $k_{B}$ is the Boltzmann constant, and $T$ is the absolute temperature of the system.
%$V_{int}$ stands for the inter-particle interaction potential and $V_{c}$ is the confinement potential.
Finally, $\vec{F}^{i}_{T}$ is
a randomly fluctuating force, which obeys the following conditions: $\langle \vec{F}_{T} \rangle = 0$
and $\langle F^{i}_{T}(t) F^{i'}_{T}(t') \rangle = 2 \eta k_{B} T \delta_{ii'} \delta(t-t')$, where $\eta$ is the friction coefficient.
Eq. (\ref{eq7}) can be written in dimensionless form as follows:
\begin{eqnarray}\label{eq8}
\frac{d\vec{r'}_{i}}{dt'} &=& D'_{i} \Gamma \Big[ - \sum_{j \neq i} \vec{\nabla '}_{i} V'_{int}(\vec{r'}_{ij}) 
\nonumber \\ 
&-& \vec{\nabla '}_{i} V'_{conf}(\vec{r'}_{i}) + \vec{F'}^{i}_{T}(t') \Big],
\end{eqnarray}
where we use the following transformation
$V_{int} = (q^{2}/\epsilon a_{0})V^{'}_{int}$, $D'_{i} = D_{i}/a^{2}_{0}$,
and introduced a coupling parameter $\Gamma$,
which is the ratio of the average potential energy to the average kinetic energy,
$\Gamma = \langle V \rangle / \langle K \rangle$,
such that $\Gamma = q^{2}/k_{B}T \epsilon a_{0}$. The time $t'$ is expressed in seconds
and distances are expressed in units of the interparticle distance $a_{0}$.
%The time $t'$ is given by $t' = Dt/a^{2}_{0}$. %$(a_{0}/q)\sqrt{m\epsilon/\kappa}$.
In what follows, we will abandon the prime ($'$) notation.
We have used a first order finite difference method (Euler method)
to integrate Eq.~(\ref{eq8}) numerically.
In the case of a straight channel, periodic boundary conditions (PBC)
were applied in the $x$-direction while in the $y$-direction the
system is confined either by a hard-wall or by a parabolic potential.
Also, we use a timestep $\Delta t = 0.001$ and the coupling parameter is set to $\Gamma = 10$.
For a circular channel, we use polar coordinates $(r,\phi)$
and model a 2D narrow channel of radius $r_{ch}$
with parabolic potential-energy profile across the channel,
i.e., in the $r$-direction.

\section{1D versus 2D diffusion in a straight channel}\label{Straight}

\subsection{Mean-square displacement (MSD) calculations}
In order to characterize the diffusion of the system, we calculate the MSD as follows:
\begin{equation}
\langle \Delta x^{2}(t) \rangle = \Big\langle \frac{1}{N} \sum_{i=1}^{N}
\left[x_{i}(t+\Delta t) - x_{i}(t)\right]^{2}  \Big\rangle_{\Delta t},
\end{equation}
where $N$ is the total number of particles and $\langle ... \rangle_{\Delta t}$ represents a time average over the time interval $\Delta t$. 
Note that in the general case (e.g., for small circular channels with the number of particles $N=20$ --- see Sec.~IV) the calculated MSD was averaged over time {\it and} over the number of ensembles~\cite{TAEA}. 
However, we found that for large $N$ (i.e., several hundred) the calculated MSD for various ensemble realizations coincide (with a maximum deviation within the thickness of the line representing the MSD). 

\begin{figure}[btp]
\begin{center}
\includegraphics*[width=7.0cm]{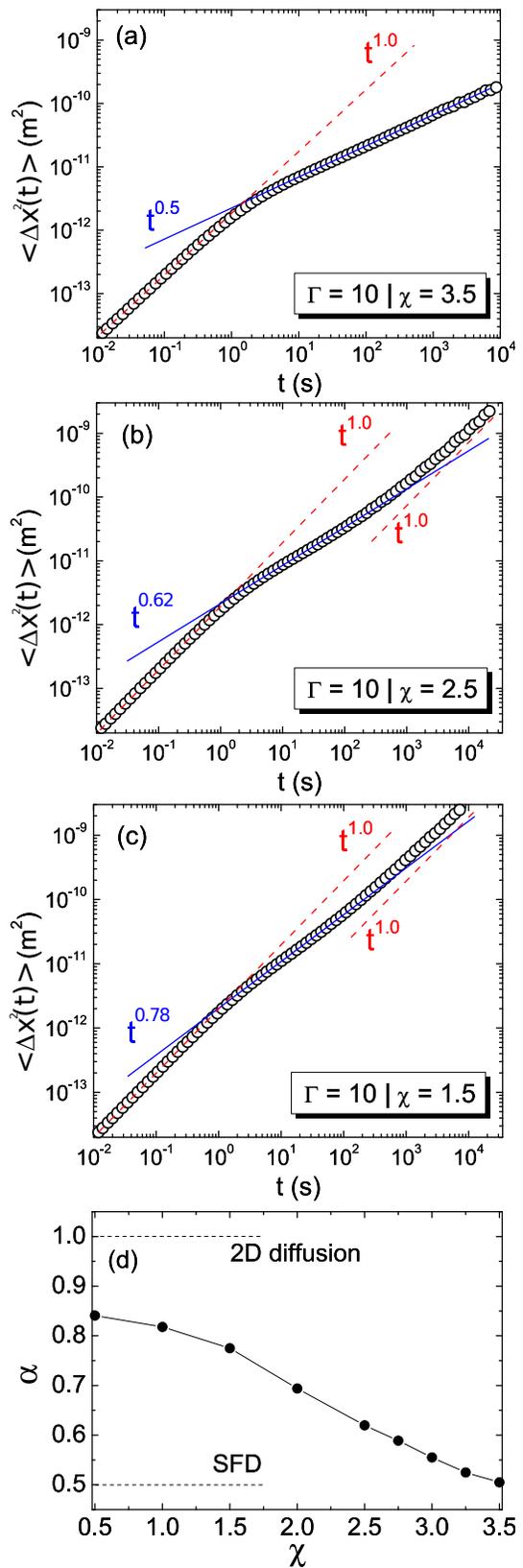}
\caption{(Color online) (a)-(c) Log-log plot of the mean-square displacement (MSD) $\langle \Delta x^{2}(t) \rangle$ as a function of time for different values of $\chi$. Different diffusion regimes can be distinguished: normal diffusion regime ($\alpha = 1.0$) and intermediate subdiffusive regime (ITR, $\alpha < 1.0$). Note that for the case of $\chi$ = 1.5, there is a normal diffusion regime (i.e. $\alpha = 1.0$) after the ITR. The dashed and solid lines in (a)-(c) are a guide to the eye. Panel (d) shows the dependence of the slope ($\alpha$) of the MSD curves (in the ITR, characterized by an apparent power-law; $\langle \Delta x^{2}(t) \rangle$ $\propto t^{\alpha}$) on the confinement strength $\chi$.}\label{MSDParabolic}
\end{center}
\end{figure}

\begin{figure}[btp]
\centering
\hspace*{-1cm}
%\begin{center}
\includegraphics*[width=7.0cm]{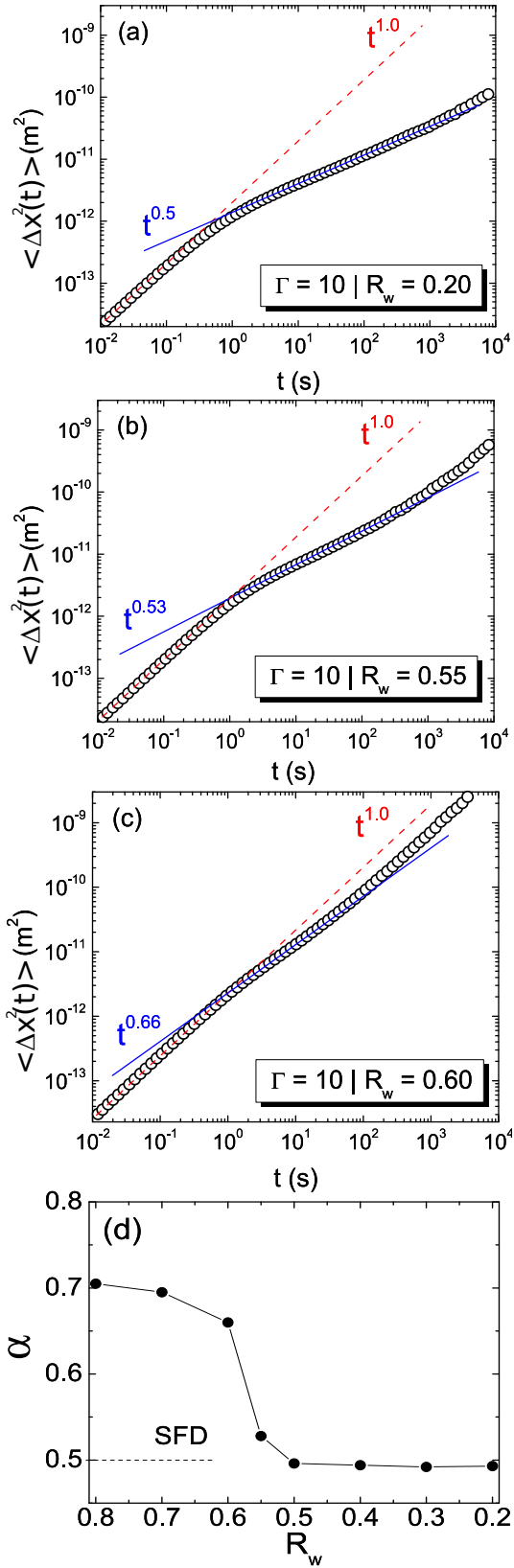}
%\end{center}
\caption{(Color online) (a)-(c) Log-log plot of the mean-square displacement (MSD) $\langle \Delta x^{2}(t) \rangle$ as a function of time for different values of $R_{w}$. Different diffusion regimes can be distinguished: normal diffusion regime ($\alpha = 1.0$) and intermediate subdiffusive regime (ITR, $\alpha < 1.0$). Note that for the case of $R_{w}$ = 0.60, there is a normal diffusion regime (i.e. $\alpha = 1.0$) after the ITR. The dashed and solid lines in (a)-(c) are a guide to the eye. Panel (d) shows the dependence of the slope ($\alpha$) of the MSD curves (in the ITR, characterized by an apparent power-law; $\langle \Delta x^{2}(t) \rangle$ $\propto t^{\alpha}$) on the confinement parameter $R_{w}$.}\label{MSDHardwall}
\end{figure}

To keep the inter-particle distance approximately
equal to unity, we defined the total number of particles $N$ for a 1D and Q1D system as
\begin{equation}
N = \frac{L}{\sqrt{1 - R_{w}^{2}}} \, \, \, ; \, \, \, R_{w} < 1,
\end{equation}
where $L$ is the size of the simulation box (in dimensionless units) in the $x$-direction. In our simulations for a straight channel geometry, we typically used $N = 400-900$ particles. We study the system for two different types of confinement potential: (i) a parabolic 1D potential in the $y$-direction, which can be tuned by the confinement strength $\chi$ and (ii) a hard-wall potential, where particles are confined by two parallel walls separated by a distance $R_{w}$.
%and $R_{w}$ is the width of the confining channel.
%For different values of the parabolic strength $\chi$ and of the width of the channel $R_{w}$,
%we obtained the MSD curves as a function of time shown in Fig.~\ref{MSDParabolic}(a)--(c) and Fig.~\ref{MSDHardwall}(a)--(c), respectively.
The results of calculations of the MSD as a function of time for different values of the confinement strength $\chi$ [Eq. (\ref{ki})] and the width of the channel $R_{w}$ are presented in Fig.\ref{MSDParabolic}(a)-(c) and Fig.\ref{MSDHardwall}(a)-(c), respectively.
%We also calculate the time evolution of the exponent $\alpha$, as in Ref.~\cite{prePZia}, using the ``double logarithmic time derivative''
%\begin{equation}\label{alphatime}
%\alpha(t) = \frac{d \, \log \, \text{MSD}}{d \, \log \, t},
%\end{equation}
%and the results are shown in Fig.~\ref{MSDParabolic}(a) and Fig.~\ref{MSDHardwall}(a).

%Log-log plot of the mean-squared displacement (MSD) curves as a function of time for different values of $\chi$ (a)--(b) and of the width $R_{w}$ (c)--(d) of the confining channel.
%Two cases are shown here:
%(a) and (c) MSD curves were calculated taking into account only the $x$-direction and
%(b) and (d) MSD are calculated taking into account both $x$- and $y$-directions.
%The red (grey) dashed lines indicate some important slopes.
%Two different regimes can be distinguished: normal diffusion regime
%and single-file diffusion (SFD) regime.

\begin{figure}[btp]
\centering
\hspace*{-1cm}
%\begin{center}
\includegraphics*[width=8.0cm]{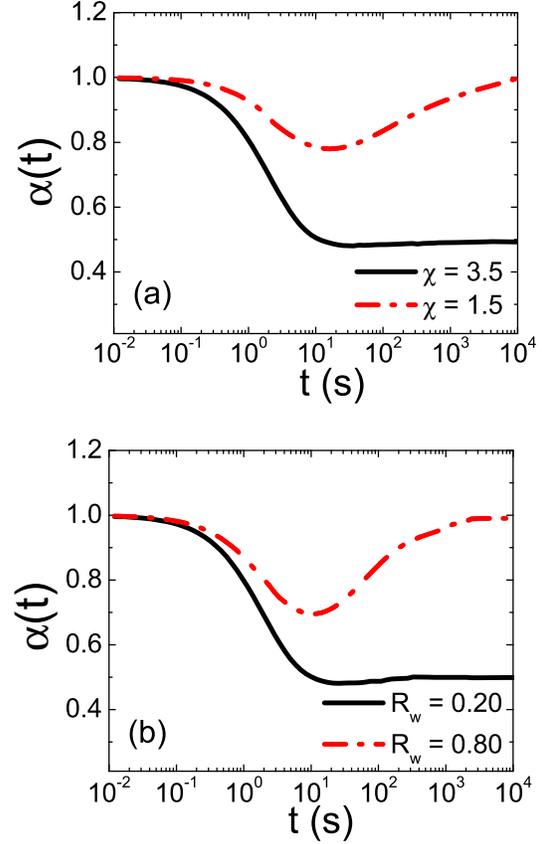}
%\end{center}
\caption{(Color online) (a)-(b) Exponent $\alpha$ as a function of time, calculated from Eq. (\ref{alphatime}) for different values of the confinement parameters $\chi$ and $R_{w}$, respectively.}\label{alphaTime}
\end{figure}

Initially, in both cases (i.e., a parabolic and a hard-wall confinement potential), the system exhibits a short-time normal diffusion behavior, where $\langle \Delta x^{2}(t) \rangle$ $\propto t^{1.0}$. This is the typical initial ``free-particle'' diffusion regime. After this initial regime, there is an intermediate subdiffusive regime (ITR). As discussed in Ref.~\cite{RKutner}, the ITR shows an apparent power-law behavior \cite{powerlaw}, where $0.5 < \alpha < 1.0$, and it was also found previously in different diffusion models \cite{PRB_28_5711,AlderAlley}. In the ITR, we found a SFD regime for either a channel with strong parabolic confinement [$\chi = 3.5$ (Fig.~\ref{MSDParabolic}(a))] or a narrow hard-wall channel [$R_{w} = 0.20$ (Fig.~\ref{MSDHardwall}(a))]. This is due to the fact that for large (small) values of $\chi$ ($R_{w}$), the confinement prevents particles from passing each other. The results for $\alpha$ in the ITR are shown as a function of $\chi$ and $R_{w}$ in Fig.~\ref{MSDParabolic}(d) and Fig.~\ref{MSDHardwall}(d), respectively. As can be seen in Fig.~\ref{MSDParabolic}(d) [Fig.~\ref{MSDHardwall}(d)], $\alpha$ increases with decreasing $\chi$ [with increasing $R_{w}$] and thus the SFD condition turns out to be broken. The values of $\alpha$ presented in these figures correspond to the minimum of the effective time dependent exponent $\alpha(t)$. Following Ref.~\cite{prePZia}, $\alpha(t)$ is calculated using the ``double logarithmic time derivative''
\begin{equation}\label{alphatime}
\alpha(t) = \frac{d \, \log \, \langle \Delta x^{2}(t) \rangle}{d \, \log \, t},
\end{equation}
and the results are shown in Fig.~\ref{alphaTime}.

The different diffusive regimes, i.e. normal diffusion regime ($\alpha = 1.0$) and SFD ($\alpha = 0.5$), were also found recently in finite-size systems \cite{SJNew,PMCentres} although the transition from SFD to normal diffusion was not analyzed. The $\alpha$-dependence on both the confinement parameters (i.e., $\alpha(\chi)$ and $\alpha(R_{w})$) presents a different qualitative behavior, namely, the SFD regime is reached after a smoother crossover in the parabolic confinement case as compared to the hardwall case. A similar smoother crossover is also found in the case of a circular channel with parabolic confinement in the radial direction. A more detailed discussion on these two different types of the behavior of $\alpha$ will be provided in Sec.~\ref{CircularChannel}.

%It is worth noting that the calculated dependence of $\alpha$ on the confinement parameter has a qualitative different shape for the case of a hard-wall potential, i.e., $\alpha(R_{w})$, as compared to the case of a parabolic potential, i.e., $\alpha(\chi)$. It is characterized by a \textit{smooth crossover} in case of a parabolic confinement [Fig.~\ref{MSDParabolic}(c)] while it is a rather \textit{sharp transition} for the hard-wall potential case [Fig.~\ref{MSDHardwall}(c)].

\subsection{``Long-time'' behavior of the MSD curves and crossing events $C(t)$}\label{LTimeMSD}

For small values of the parabolic confinement (e.g., $\chi = 1.5$), the MSD curves present three different diffusive regimes: (i) a short-time normal diffusion regime, where MSD $\langle \Delta x^{2}(t) \rangle$ $\propto t^{1.0}$; (ii) a subdiffusive regime with $\langle \Delta x^{2}(t) \rangle$ $\propto t^{\alpha}$, where $0.5 < \alpha < 1.0$ and (iii) a ``long-time'' diffusion regime, which is characterized by $\langle \Delta x^{2}(t) \rangle$ $\propto t^{1.0}$. Note that the ``long-time'' term used here is not to be confused with the long-time used for \textit{infinite} systems, as discussed in the Introduction. However, for large values of the parabolic confinement (e.g., $\chi = 3.5$), we observe only two distinct diffusive regimes, namely: (i) a short-time normal diffusion regime ($\langle \Delta x^{2}(t) \rangle$ $\propto t^{1.0}$) and (ii) a SFD regime (i.e., $\langle \Delta x^{2}(t) \rangle$ $\propto t^{0.5}$).

\begin{figure}[btp]
\centering
\hspace*{-1cm}
%\begin{center}
\includegraphics*[width=8.0cm]{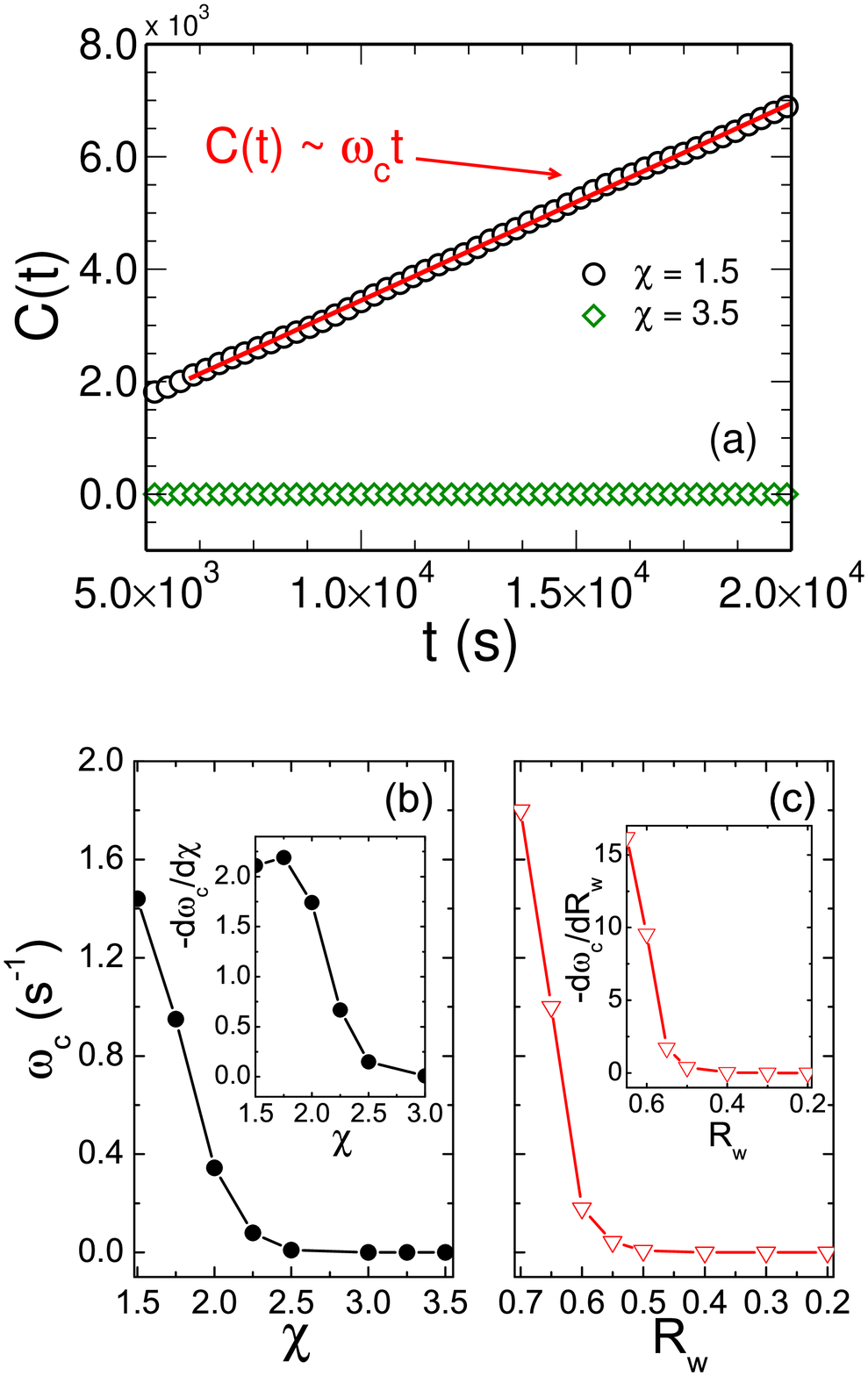}
%\end{center}
\caption{(Color online) (a) Number of crossing events $C(t)$ as a function of time for $N = 400$ particles, for $\chi = 1.5$ (black open circles) and $\chi = 3.5$ (green open diamonds). The solid red line is a linear fit to $C(t)$. Panels (b) and (c) show the rate of the crossing events $\omega_c$ as a function of the confinement potential parameters ($\chi$ and $R_{w}$).
The insets in the panels (b) and (c) show the derivatives, 
$d\omega_{c}(\chi)/d \chi$ and $d\omega_{c}(R_w)/d R_w $, 
correspondingly. 
}
\label{nCross}
\end{figure}

One question that arises naturally is whether this normal diffusion regime (i.e., $\langle \Delta x^{2}(t) \rangle$ $\propto t^{1.0}$ for ``long-times'') is an effect of the \textit{colletive} motion of the system (center-of-mass motion) or an effect of the single-particle jumping process, since the confinement potential $\chi = 1.5$ allows particles bypass. 
In order to answer this question, we calculate the number of crossing events $C(t)$ as a function of time and results are shown in Fig.~\ref{nCross}(a).
We found that for small values of the confinement potential (e.g., $\chi = 1.5$) the number of crossing events grows linearly in time, i.e., $C(t) \propto \omega_{c}t$, where $\omega_{c}$ is the rate of crossing events.
On the other hand, a strong confinement potential (e.g., $\chi = 3.5$) prevents particles from bypassing, and thus $C(t) = 0$ during the whole simulation time. 

Therefore, the ``long-time'' normal diffusive behavior (i.e., $\langle \Delta x^{2}(t) \rangle$ $\propto t^{1.0}$ for ``long-times'') found in our simulations for the case where the SF (single-file) condition is broken (e.g., $\chi = 1.5$) is {\it not} due to a collective  (center-of-mass) diffusion. 
Instead, this normal diffusive behaviour is due to a single-particle jumping process, which happens with a constant rate $\omega_{c} > 0$ for the case of small values of the confinement ($\chi = 1.5$) and $\omega_{c} = 0$ (for $\chi = 3.5$). The same analysis was done for the case of the hard-wall confinement potential, and the results are found to be the same as for the parabolic confinement.

Nevertheless, we point out that the collective diffusion does indeed exist, but our results from simulations do not allow us to observe this collective (center-of-mass) diffusion regime because of the large size of our chain of particles ($N = 400 - 900$). Simulations with $N = 80 - 100$, and excluding the possibility of mutual bypass (strong confinement potential), allowed us to observe that the $\langle \Delta x^{2}(t) \rangle$ $\propto t^{1.0}$ regime is recovered in the ``long-time'' limit. In Sec.~V, we will further discuss the long-time limit using a model of discrete sites.

%%%  

As we demonstrated above, the transition from pure 1D diffusion (SFD) characterized by $\alpha = 0.5$ to a quasi-1D behavior (with $\alpha > 0.5$) could be either more ``smooth'' (as in Fig.~1(d), for a parabolic confinement) or more ``abrupt'' (as in Fig.~2(d), for a hard-wall confinement). One can intuitively expect that this difference in behavior can manifest itself also in the crossing events rate $\omega_{c}$, i.e., that $\omega_{c}$ as a function of $\chi$ (or $R_w$) should display a clear signature of either ``smooth'' or ``abrupt'' behavior. 
 
However, the link between the two quantities, i.e., the exponent, $\alpha(\chi/R_w)$, and the crossing events rate, $\omega_{c}(\chi/R_w)$ is not that straightforward. To understand this, let us refer to the long-time limit (which will be addressed in detail within the discrete-site model in Sec.~V). As we show, in the long-time limit the exponent $\alpha$ is defined by one of the two conditions: $\omega_{c} = 0$ (then $\alpha = 0.5$) or $\omega_{c} > 0$ (then $\alpha = 1$) and it does {\it not} depend on the specific value of $\omega_{c}$ provided it is nonzero. Therefore, in the long-time limit the transition between 1D to 2D behavior {\it is not sensitive} to the particular behavior of the function $\omega_{c}(\chi/R_w)$. 
 
Although for ``intermediate'' times (considered in this section) the condition $\omega_{c} = 0$ {\it or} $\omega_c > 0$ is not critical, nevertheless, very small change in the crossing events rate $\omega_{c}(\chi/R_w)$ strongly influences the behavior of the exponent $\alpha(\chi/R_w)$. This is illustrated in Figs.~4(b, c). In Fig.~4(b), the function $\omega_{c}(\chi)$ gradually decreases from 1.45 to 0 for $\chi$ varying in a {\it broad} interval from 1.5 to 3 (note that the segment of $\omega_{c}(\chi)$ for $2.5 < \chi < 3$ is nonzero which can be seen in the inset of Fig.~4(b) showing the derivative $d\omega_{c}(\chi)/d \chi$). Correspondingly, the transition from $\alpha = 0.5$ to $\alpha \approx 0.8$ in that interval of $\chi$ is ``smooth'' (see Fig. 1(d)). On the other hand, the function $\omega_{c}(R_w)$ shown in Fig.~4(c) mainly changes (note the change of the slope $d\omega_{c}(R_w)/d R_w $ shown in the inset of Fig.~4(c)) in a {\it narrow} interval $0.5 < R_w < 0.6$. Respectively, the transition for the function $\alpha(R_w)$ occurs in the narrow interval $0.5 < R_w < 0.6$ and thus is (more) ``abrupt''. 
 
%%%

%

%In order to understand this reentrant behavior, we have calculated the mean distance $\langle |y| \rangle$ of the particles from the center of the channel and the standard deviation $\sigma$, respectively, using the probability distribution $P(y)$, as
%\begin{eqnarray}
%\langle |y| \rangle &=& \frac{\sum_{i}^{N_{s}} |y_{i}| P_{i}(y_{i}) }{\sum_{i}^{N_{s}}P_{i}(y_{i})}; \\
%\sigma &=& \left( \frac{1}{N_{s}} \sum_{i=1}^{N_{s}} (|y_{i}| - \langle |y| \rangle )^{2}  \right)^{1/2}
%\end{eqnarray}
%where $P(y)$ is the probability of finding a particle between $y$ and $y + \Delta y$, and $N_{s}$ is the total number of data points.
\subsection{Distribution of particles along the $y$-direction}
\begin{figure}[btp]
%\vspace*{-1cm}
\vspace*{0.5cm}
\hspace*{-3cm}
\begin{center}
\includegraphics*[width=8.0cm]{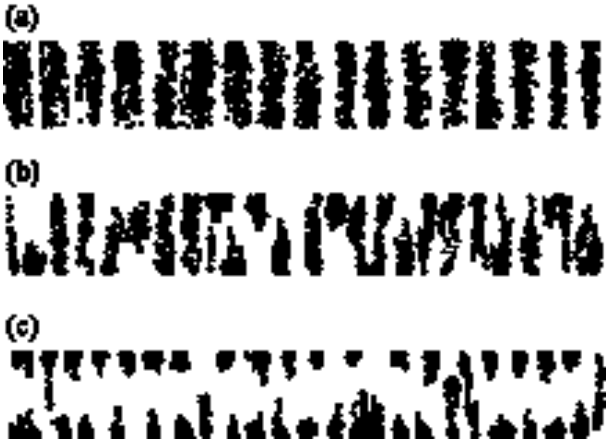}
\end{center}
\caption{For the hard-wall confinement case, we show typical trajectories of particles (i.e. 10$^{6}$ MD simulation steps) confined by the channel of width (a) $R_{w} = 0.20$, (b) $R_{w} = 0.60$ and (c) $R_{w} = 0.80$.}\label{HardSnap}
\end{figure}

For the ideal 1D case, particles are located on a straight line. Increasing the width $R_{w}$ of the confining channel will lead to a zig-zag transition \cite{gioPRE, gioPRB}. This zig-zag configuration can be seen as a distorted triangular configuration in this transition zone. Further increase of $R_{w}$ brings the system into the 2D regime, where the normal diffusion behavior is recovered (see Fig.~\ref{HardSnap}). 

\begin{figure}[btp]
%\vspace*{-1cm} \hspace*{-3cm}
\begin{center}
\includegraphics*[width=8.0cm]{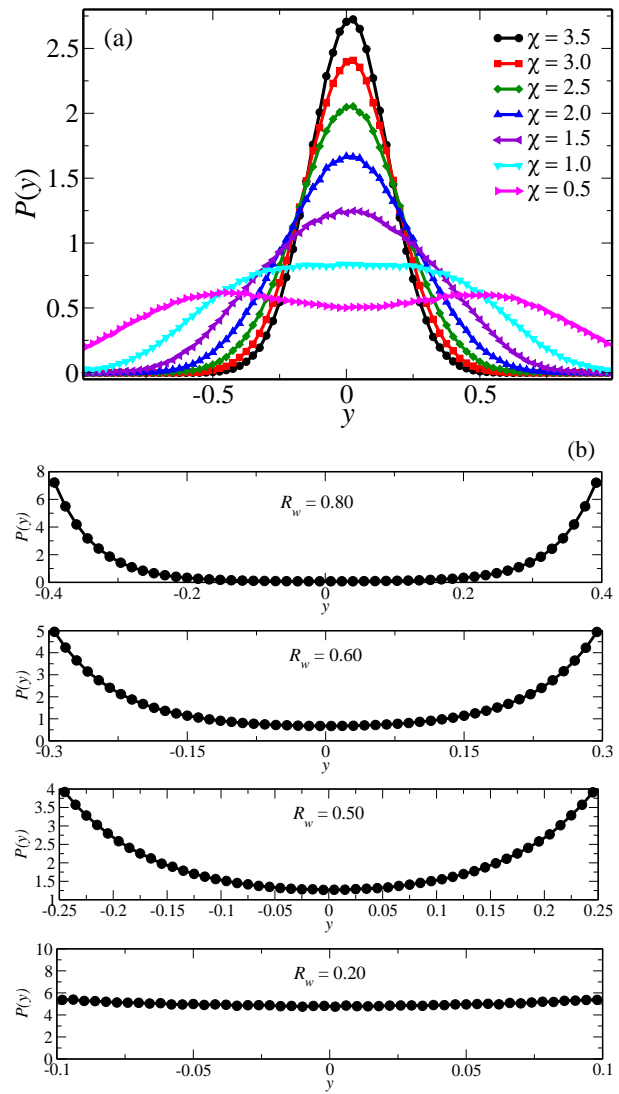}
\end{center}
\caption{(Color online) Probability distribution of the particle density $P(y)$ along the $y$-direction are shown for
(a) different values of $\chi$ (parabolic 1D confinement)
and (b) four different values of the width $R_{w}$ of the channel (hard-wall confinement).}\label{DistPy}
\end{figure}

For the parabolic 1D confinement, we can see [Fig.~\ref{DistPy}(a)] that the distribution of particles $P(y)$ along the channel is symmetric along the axis $y = 0$. Also, for large values of $\chi$ (e.g., $\chi = 3.5$) particles are confined in the $y$-direction and thus can move only in the $x$-direction, forming a single-chain structure. As the confinement decreases ($\chi \rightarrow 0$), the distribution of particles $P(y)$ broadens resulting in the crossover from the SFD regime ($\chi = 3.5$) to the 2D normal diffusion regime ($\chi = 0.5$). Note that for small values of $\chi$ (e.g., $\chi = 0.5$), the system forms a two-chain structure (represented by two small peaks of $P(y)$ in Fig.~\ref{DistPy}(a)), thus allowing particles to pass each other.

\section{Diffusion in a circular channel}\label{CircularChannel}

In the previous section, we analyzed the transition (crossover)
from the SFD regime to 2D diffusion in narrow channels of increasing
width.
The analysis was performed for a straight channel  with either hard-wall
or parabolic confinement potential.
However, in terms of possible experimental verification of the
studied effect, one faces an obvious limitation of this model:
although easy in simulation, it is hard to experimentally fulfill
the periodic boundary conditions at the ends of an {\it open} channel.
Therefore, in order to avoid this difficulty, in SFD experiments
%with colloids \cite{weiScience} and metallic charged balls \cite{MSJ}
\cite{weiScience,MSJ} circular channels were used.

In this section, we investigate the transition (crossover) from SFD to 2D-diffusion in a system of interacting particles diffusing in a channel of \textit{circular} shape. In particular, we will study the influence of the strength of the confinement (i.e., the depth of the potential profile across the channel) on the diffusive behavior.
%
%(Note that changing the strength of the parabolic potential is
%equivalent to changing the width of the channel.)
%
Without loss of generality, we will adhere to the specific conditions
and parameters of the experimental set-up used in Ref.~\cite{MSJ}.
An additional advantage of this model is that the motion of the system of charged metallic balls \cite{MSJ} is {\it not} overdamped, and we will solve the full Langevin equations of motion to study the diffusive
behavior of the system.

We consider $N$ particles, interacting through a Yukawa potential [Eq.~(\ref{eq3})],
which are embedded in a ring channel of radius $r_{ch}$.
We define a parabolic confinement potential across the channel in the form (\ref{vc})
%\begin{equation}\label{eq9}
%V_{conf}(r_i)=\beta\left(r_{ch}-r_i\right)^2,
%\end{equation}
where parameter $\beta$ is chosen as follows:
\begin{equation}
%\label{eq10}
\beta=\frac{V_0}{\gamma r_0^2},\ \
V_0=\frac{q^2}{\epsilon}\sum_{i\neq j}
\frac{\exp \left[-2 \kappa r_{ch}
\sin\left(\frac{\phi_i-\phi_j}{2}\right)\right]}{2r_{ch}
\sin\left(\frac{\phi_i-\phi_j}{2}\right)},
\end{equation}
when all the particles are equidistantly distributed
along the bottom of the circular channel.
It should be noted that in this case, $V_{0}$ is approximately
equal to $V_{gs}$ due to the weak Yukawa interaction, which slightly
shifts the particles away from the bottom of the channel.
Such a choice of $V_{0}$ is related to the fact that
we study the influence of the confinement on the diffusion
and, therefore, the potential energy of the particles must
be of the order of the inter-particle interaction energy.
Parameter $r_0$
%$r_0 = r_{ch}/22.5$
characterizes the distance where the external potential reaches the value
$V_0/\gamma$, and $V_{gs}$ is the energy of
the ground state of the system of $N$ particles as
defined by Eq.~(\ref{eq4}).
Parameter $\gamma$ plays the role of a control parameter.
By changing $\gamma$ we can manipulate the strength of the
confinement and, therefore, control the fulfillment
of the single-file condition.
Increase in $\gamma$ corresponds to a decrease in the depth
of the confinement (\ref{vc}) which leads to the
expansion of the area of radial localization of particles.
Therefore, an increase of $\gamma$ results in a similar effect
(i.e., spatial delocalization of particles) as an increase
of temperature, i.e., parameter $\gamma$ can be considered
as an ``effective temperature''.
Note that such a choice of the parameter that controls the
confinement strength is rather realistic.
In the experiment of Ref.~\cite{MSJ} with metallic balls,
the parabolic confinement was created by an external electric field,
and the depth of the potential was controlled by tuning the strength
of the field.

To study diffusion of charged metallic balls, we solve
the Langevin equation of motion in the general form
(i.e., with the inertial term $\propto m$),
\begin{eqnarray}
\label{eq11}
m \frac{d^2 \vec{r}_i}{dt^2} &=& - \eta \frac{d\vec{r}_i}{dt} - \sum_{j,i\neq j} \vec{\nabla}V_{int} (\vec{r}_{ij}) 
\nonumber \\ 
&-& \vec{\nabla}V_{c}(\vec{r}_{i}) + \vec{F}^{i}_{T},
\end{eqnarray}
where $m = 2.5 \times 10^{-6}$~kg \cite{MSJ} is the mass of a particle, $\eta$ is the friction coefficient (inverse to the mobility). 
Here all the parameters of the system were chosen following the
experiment \cite{MSJ}, and $\lambda_{D} = 4.8 \times 10^{-4}~m$, $\Gamma=1$
(which is a typical experimental value, see, e.g., also \cite{weiScience}).
Correspondingly, mass is measured in kg, length in m, and
time in seconds. 
Also, following Ref.~\cite{MSJ}, we took a channel of radius $r_{ch} = 9$~mm (in the experiment~\cite{MSJ}, the external radius of the channel was 10~mm, and the channel width 2~mm; note that in our model we do not define the channel width: the motion of a particle in the transverse direction is only restricted by the parabolic confinement potential). We also took experimentally relevant number of diffusing particles, $N$, varying from $N=12$ to $N=40$ (in the experiment~\cite{MSJ}, the ring channel contained $N=12$ or $N=16$ diffusing balls).

\begin{figure}[btp]
\begin{center}
\includegraphics*[width=8.5cm]{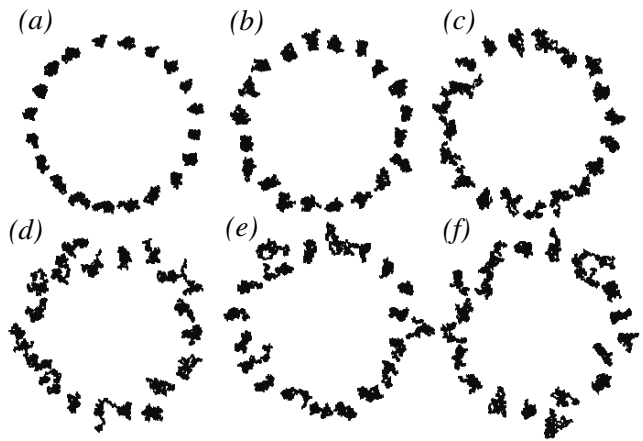}
\end{center}
\caption{
Trajectories of $N=20$ particles diffusing in a ring of radius $r_{ch}=9$~mm for $10^{6}$ consequent time steps for different values of $\gamma$. $\gamma$=1 (a), 2 (b), 3 (c), 5 (d), 7 (e), 9 (f).
}
\label{FigSnapshots}
\end{figure}

Fig.~\ref{FigSnapshots} shows
the results of calculations of the trajectories of 
$N=20$ particles diffusing in a ring of radius $r_{ch}=9$~mm 
for the first 10$^{6}$ MD steps for various values
of the parameter $\gamma$.
As can be seen from the presented snapshots, the radial
localization of particles weakens with increasing $\gamma$.
At a certain value of $\gamma$ this leads to the breakdown of
the single-file behavior (Figs.~\ref{FigSnapshots}(c)-(f)).

%The single-file condition is fulfilled for $\gamma\leq2$ (a), (b), while for $\gamma>2$ (c) to (f) the SF condition is broken.

\subsection{Breakdown of SFD}

\begin{figure}[btp]
\begin{center}
\includegraphics*[width=8.5cm]{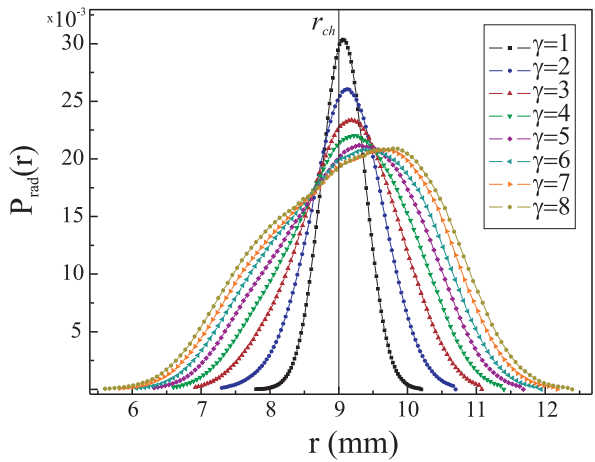}
\end{center}
\caption{
(Color online) The distribution of the probability
density of particles $P_{\text{rad}}(r)$ 
in a circular channel of radius $r_{ch}=9$~mm 
along the radial direction $r$.
The different curves correspond to various $\gamma$.
Increasing $\gamma$ the width of the distribution
$P_{\text{rad}}(r)$ increases due to a weakening of the confinement.
}
\label{FigProb}
\end{figure}

It is convenient to introduce the distribution of the
probability density of particles in the channel $P_{\text{rad}}$
along the radial direction $r$.
In order to calculate the function $P_{\text{rad}}(r)$ we divided
the circular channel in a number of coaxial thin rings.
The ratio of the number of observations of particles in
a sector of radius $r_i$ to the total number of observations
during the simulation is defined as the probability
density $P_{\text{rad}}(r_i)$.
In Fig.~\ref{FigProb}, the probability density $P_{\text{rad}}(r)$
is presented for different values of $\gamma$.
%All the distributions are approximately centered near $r_{ch}$.
With increasing $\gamma$, the distribution of the
probability density $P_{\text{rad}}(r)$ broaden and the maximum
of the function $P_{\text{rad}}(r)$ shifts away from the center of
the channel (see Fig.~\ref{FigProb}).
The latter is explained by the softening of the localization
of particles with increasing $\gamma$, which tend to occupy
an area with a larger radius due to the repulsive inter-particle
interaction.
Simultaneously, the distribution of the probability density
$P_{\text{rad}}(r)$ acquires an additional bump indicating the
nucleation of a two-channel particle distribution \cite{nonzigzag}.
The observed broadening and deformation of the function
$P_{\text{rad}}(r)$ is indicative of a gradual increase of the probability
of mutual bypass of particles (i.e., the violation of the SF (single-file) condition,
also called the ``overtake probability'' \cite{ambjornsson})
with increasing $\gamma$.

\begin{figure}[btp]
\begin{center}
\includegraphics*[width=7.5cm]{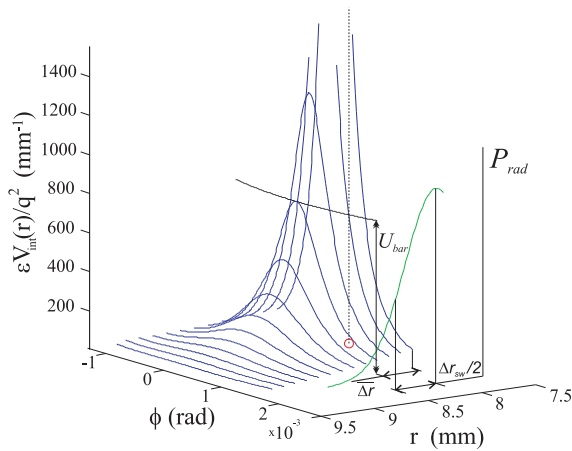}
\end{center}
\caption{
(Color online) Spatial distribution of the potential
$V_{int}(r,\phi)$ created by a particle (red (grey) circle)
and the qualitative distribution of the probability
density of particles in circular channel $P_{\text{rad}}(r)$
(green (light grey) line) along the radial direction $r$.
The function $\overline{\Delta r}$ determines an approximate
radial distance between particles when the potential barrier
$U_{bar}$ becomes ``permeable'' for given temperature $T$.
The function $\Delta r_{sw}$ characterizes a width of the
distribution $P_{\text{rad}}(r)$ at this temperature $T$.
}
\label{Fig7}
\end{figure}

Let us now discuss a qualitative criterion for the breakdown
of SFD, i.e., when the {\it majority} of particles leave
the SFD mode.
For this purpose, let us consider a particle
in the potential created by its close neighbor
(which is justified in case of short-range Yukawa interparticle
interaction and low density of particles in a channel)
shown in Fig.~\ref{Fig7}.
Different lines show the interparticle potential $V_{int}$
as a function of angle $\phi$ for different radii $r$.
For small values of $\gamma$, the center of the distribution 
$P_{\text{rad}}(r)$ (see Fig.~7) almost coincides with the center of the channel (i.e., with the minimum of the confinement potential profile)  and the distribution $P_{\text{rad}}(r)$ is narrow. 
Therefore, mutual passage of particles is impossible,
i.e., the SF condition is fulfilled.
The asymmetric broadening of the function $P_{\text{rad}}(r)$ with
increasing $\gamma$ results in an increasing probability of
mutual bypass of particles which have to overcome a barrier
$U_{bar}$ (see Fig.~\ref{Fig7}).
This becomes possible when $U_{bar} \lesssim k_{B}T$.
In other words, the thermal energy $k_{B}T$ determines some minimal
width $\overline{\Delta r}$ between adjacent particles when
the breakdown of the SF condition becomes possible.

It is clear that ``massive'' violation of the SF condition
(i.e., when the majority of particles bypass each other) occurs
when the halfwidth $\Delta r_{sw}$ of the distribution of the
probability density $P_{\text{rad}}(r)$ obeys the condition:
\begin{equation}
\label{nsf01}
\Delta r_{sw} \gtrsim \overline{\Delta r}.
\end{equation}
The function $\Delta r_{sw}$ is defined by the ratio of
the thermal energy $k_{B}T$ to the external potential $U_{conf}(r)$
and is of the same order as $\widetilde{\Delta r}$:
\begin{equation}
\frac{V_0}{\gamma r_0^2}\cdot (\widetilde{\Delta r}/2)^2\approx k_{B}T.
\end{equation}
Therefore the criterion (\ref{nsf01}) can be presented in the form:
\begin{equation}
\Delta r_{sw}\approx\widetilde{\Delta r} \gtrsim \overline{\Delta r}.
\end{equation}

This qualitative analysis of the breakdown of the SFD regime 
clarifies the role of the width and the shape of the distribution 
of the probability density influenced by the asymmetry of 
the circular channel.

\subsection{Diffusion regimes}

The MSD $\langle \Delta\phi^2(t) \rangle$ is calculated
as a function of time $t$ as:
\begin{equation}\label{eq10}
\left\langle\Delta\phi^2(t)\right\rangle
=\left\langle \frac{1}{N_{\text{par}}N_{\text{ens}}}\sum_{i,j} \left[\Delta\phi_{ij}(\tau+t) - \Delta\phi_{ij}(t)\right]^2 \right\rangle_t,
\end{equation}
where $N_{\text{par}}$ is the total number of particles of an ensemble and $N_{\text{ens}}$ is the total number of ensembles. In our calculations, the number of ensembles was chosen 100 for a system consisting of 20 particles.

%The MSD $\langle \Delta\phi^2 \rangle$ is calculated as a function of time $t$:
%\begin{equation}\label{eq10}
%\leftr\langle\left\langle\Delta\phi^2\right\rangle_{ens}\right\rangle_{par}
%=\frac{1}{N_{par}N_{ens}}\sum_{i}\sum_{j}
%\left( \left\langle\Delta\phi_{i,j}^2\right\rangle_t -
%\left\langle\Delta\phi_{i,j}\right\rangle_t^2 \right),
%\end{equation}
%where $\langle...\rangle_{par}$ denotes averaging over all particles of a given ensemble, and $\langle..\rangle_{ens}$ denotes averaging over various ensembles. In our calculations, the number of ensembles was chosen 100 for a system consisting of 20 particles.

\begin{figure}[!btp]
\begin{center}
\includegraphics*[width=7.0cm]{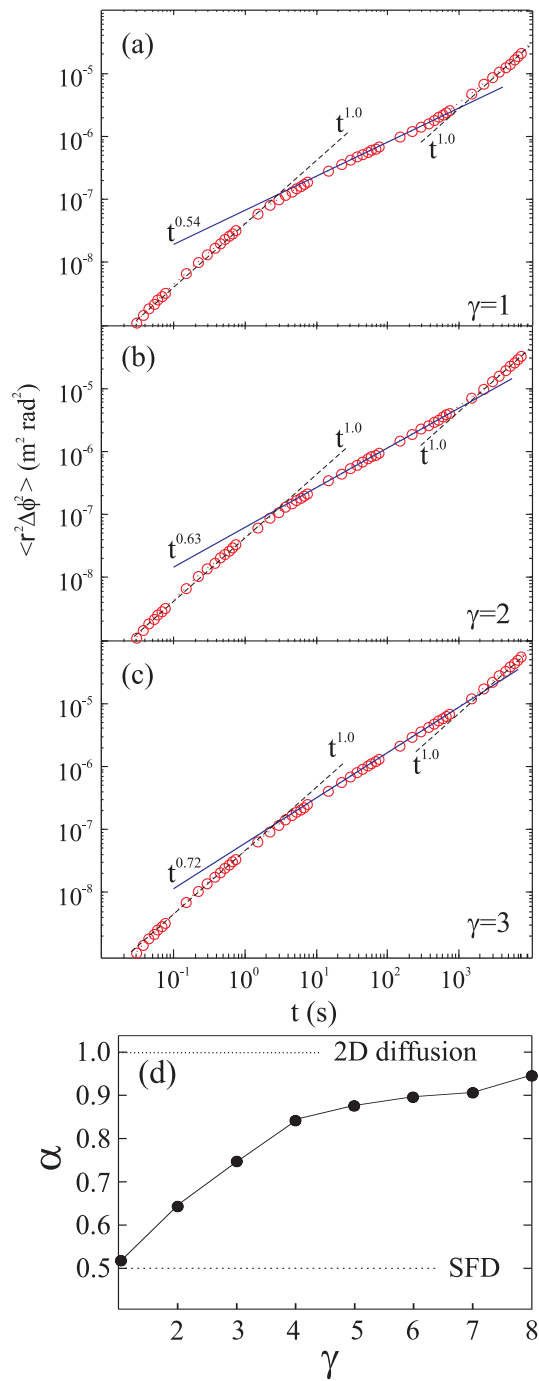}
\end{center}
\caption{
(Color online)
(a)--(c): Log-log plot of the mean-squared displacement (MSD) $\langle \Delta\phi^2 \rangle$ as a function of time
for different values of the ``effective'' temperature $\gamma$ = (a) 1, (b) 2, and (c) 3.
Here $(N_{\text{ens}}=100, N_{\text{par}}=20)$.
(d) The diffusion exponent $\alpha$ as a function of $\gamma$.
Increase of the ``effective'' temperature $\gamma$ leads to the
gradual transformation of the single-file regime of diffusion into
the diffusion regime of free particles.}
\label{FigSFDring}
\end{figure}

The time dependence of the MSD for different values of $\gamma$ is
shown in Fig.~\ref{FigSFDring}(a)--(c).
%***
Initially the system exhibits normal diffusion, where $\langle \Delta\phi^2 \rangle$ $\propto t^{1.0}$. This regime is followed by an intermediate subdiffusive regime, where the $\langle \Delta\phi^2 \rangle$ $\propto t^{\alpha}$ ($0.5 < \alpha < 1.0$). For longer times, the system recovers ``long-time'' normal diffusion (see discussions in Sec. IIIB), with $\langle \Delta\phi^2 \rangle$ $\propto t^{1.0}$. As in the case of straight channel geometry, this second crossover (i.e., from intermediate subdiffusion to ``long-time'' normal diffusion) can also be due to two other reasons: (i) due to a collective (center-of-mass) diffusion or (ii) due to a single-particle jumping process. However, for the simulations in the case of a circular geometry, the number of particles is relatively small (taking the fact that this is a finite-size system), and therefore, the crossover from sublinear to linear regime is due to a collective (center-of-mass) diffusion. We further address this issue in Sec.~V, where we consider a discrete 
site model and we exclude the center-of-mass motion. 
%***

%It can be seen that for small $\gamma$ the normal diffusion regime, i.e. $\langle \Delta\phi^2 \rangle$ $\propto t^{1.0}$ is followed by a regime of slowing down of diffusion, which in turn is followed by an acceleration of diffusion \cite{submitEPL}.

%***
%For $\gamma > 1$ the SFD region quickly shrinks $(\gamma=2)$, and then completely disappears $(\gamma>2)$, being replaced by yet a subdiffusive behavior where $\langle \Delta\phi^2 \rangle$ $\propto t^{\alpha}$ ($0.5< \alpha < 1.0$).
%***

Fig.~\ref{FigSFDring}(d) shows $\alpha$ as a function of $\gamma$.
The function $\alpha(\gamma)$ experiences a monotonic
gradual crossover from the $\alpha = 0.5$ to a
$\alpha \lesssim 1$-regime.
Note that the observed deviation from the normal diffusion behavior
for large $\gamma$ (Fig.~\ref{FigSFDring}) is related
to the presence of, though weak but nonzero, external confinement
in the radial direction.
This change of the diffusive behavior is explained by a weakening
of the average radial localization of particles with increase of
$\gamma$ (Fig.~\ref{FigProb}) and, as a consequence, by an increase
of the probability of mutual bypass of particles.

%The MSD-curves which take into account diffusion in both the longitudinal (angular)
%and transverse (radial) directions
%$\langle\Delta R^2\rangle=\langle r^2\Delta\phi^2\rangle+\langle\Delta r^2\rangle$
%are presented in Fig.~\ref{FigSFDring}(b) for different $\gamma$. As it is seen from
%Fig.~\ref{FigSFDring}(b) all the regimes are reproduced (i.e., the normal diffusion
%regime and the SFD regime) which are present in Fig.~\ref{FigSFDring}(a). However
%a similar behavior as in the case of a straight channel with parabolic confinement
%(see Fig.~1(b)), namely, the subdiffusive regime ($\alpha<0.5$) is observed.
%In other words, the qualitative modification of the diffusion curves is the same as
%in the case of a straight channel.

The observed crossover between the 1D single-file and 2D
diffusive regimes, i.e., $\alpha(\gamma)$-dependence, shows a
significant different qualitative behavior as compared to the case of a hard-wall
confinement potential considered in
Sec.~\ref{Straight}, where a rather sharp transition between the two
regimes was found [Fig.~\ref{MSDHardwall}(d)].
The different behavior is due to the different
confinement profiles and can be understood from the analysis
of the distribution of the probability density of particles for
these two cases.
In the case of a hard-wall channel, the uncompensated (i.e., by
the confinement) interparticle repulsion leads to a higher
particle density near the boundaries rather than near the center
of the channel (see Fig.~\ref{HardSnap} and Fig.~\ref{DistPy}(b)).
As a consequence, the breakdown of the SF condition --- with
increasing width of the channel --- happens simultaneously
for {\it many} particles in the vicinity of the boundary
resulting in a sharp transition (see Fig.~\ref{MSDHardwall}(d)).
On the contrary, in the case of parabolic confinement, the
density distribution function has a maximum --- sharp or broad,
depending on the confinement strength --- near the center of
the channel (see Figs.~\ref{DistPy}(a) and \ref{FigProb}).
With increasing the ``width'' of the channel (i.e., weakening
its strength), only a small {\it fraction} of particles undergoes
the breakdown of the SF condition.
This fraction gradually increases with decreasing strength
of the confinement, therefore resulting in a smooth crossover
between the two diffusion regimes.

\section{Discrete site model: The long-time limit}

The calculated MSD for different geometries and confinement potentials
allowed us to explain the evolution of the subdiffusive regime with varying
width of the channel (or potential strength in case of a parabolic potential).
However, the obtained results are only valid for the intermediate regime
and therefore they only describe the ``onset'' of the long-time behavior.
The problem of accessing the long-time behavior in a finite chain is related to
the fact that sooner or later (i.e., depending on the chain length) the interacting system
will evolve into a collective, or ``single-particle'', diffusion mode which is
characterized by $\alpha = 1.0$.
Thus the question is whether the observed behavior holds for the long-time limit,
i.e., is the transition from $t^{0.5}$ to $t^{1.0}$ behavior smooth?

To answer this question, we considered a simple model,
i.e., a linear discrete chain of fixed sites filled with either particles
or ``holes'' (i.e., sites not occupied by particles)
(for details, see Ref.~\cite{PRB_28_5711}; 
this model was also recently used in Ref.~\cite{SSPRE2006}). 
The particles can move along the
chain only due to the exchange with adjacent vacancies (i.e., with holes).
Within this model, the long-time diffusion behavior was described {\it analytically}
for an infinite linear chain as well as for a finite cyclic chain~\cite{PRB_28_5711}.
In particular, this model predicts that:
(i) If the chain is infinite then the long-time power law of the diffusion
curve $\alpha$ is $0.5$ (i.e., MSD $\langle \Delta x^{2}(t) \rangle$ $\propto t^{0.5}$);
(ii) If the chain is finite then the subdiffusive regime with $\alpha=0.5$
is followed by
either $\alpha=1.0$ regime (if the cyclic boundary condition is realized),
or by $\alpha=0$ regime, i.e., the regime of saturation
(if no cyclic boundary condition is imposed~\cite{PRE80-051103}).
The latter regime is reached for times longer than the ``diffusion time''
of a ``hole'' along the whole chain $t_{\text{chain}}$.

Let us now apply this model to a finite-size chain of particles.
For this purpose, we assume that adjacent particles are able to exchange their
positions with some probability $P$ at every time step.
For example, probability $P=0.1$ means that a couple of any adjacent particles
certainly exchange their positions once for every $10$ time steps.

\begin{figure}[!ht]
\begin{center}
\includegraphics*[width=7.5cm]{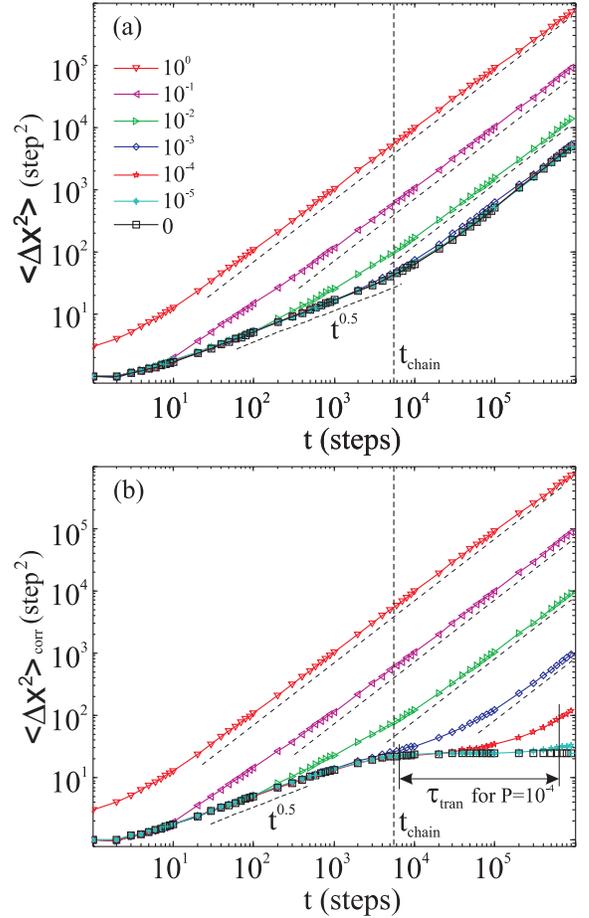}
\end{center}
\caption{(Color online)
Log-log plot of the MSD $\langle \Delta x^{2}(t) \rangle$ (a) and corrected MSD 
$\langle \Delta x^{2}(t) \rangle$$_{\text{corr}}$ (b)
as a function of time for different values of the probability $P$ of bypassing.
Averaging was done over $N_{\text{sim}}=1000$ ensembles.}
\label{FigSFDSC}
\end{figure}

The results of our calculations of the MSD performed using this model are presented
in Fig.~\ref{FigSFDSC}(a).
We used the following parameters: the chain length is $N_s=150$
sites and $N_h=1$ hole.
Averaging was done over $1000$ ensembles.
The calculation was performed for the following values of the probability:
$P=0, 10^{-5}, 10^{-4}, 10^{-3}, 0.01, 0.1$, and $1$.

We see in Fig.~\ref{FigSFDSC}(a) clearly the above-mentioned two diffusion regimes, i.e.,
with the MSD $\langle \Delta x^{2}(t) \rangle$ $\propto t^{0.5}$ and $\propto t^{1.0}$.
The characteristic time $t_{\text{chain}}$ shifts towards lower values with increasing $P$.
However this analysis (Fig.~\ref{FigSFDSC}(a)) does not allow to distinguish
the contributions to the long-time behavior ($\propto t^{1.0}$) due to:
(i) the breakdown of single-file condition (i.e., diffusion due to particle exchanges),
and (ii) the ``collective'' diffusion (chain ``rotation'').

To overcome this difficulty, we exclude the ``collective'' diffusion of the system
and introduce a modified MSD $\langle \Delta x^{2}(t) \rangle$$_{\text{corr}}$ 
(which is so-called ``roughness'' of the system of particles, as discussed in Ref.~\cite{PMCentres}) 
as follows:
$$\langle \Delta x^2 \rangle_{\text{corr}}= \langle (x-\bar{x})^2 \rangle ,$$ where $\langle ... \rangle$ is the average over time; $\bar{x}$ is the average of an ensemble of particles at a given time, or ``collective'' coordinate. It should be noted that $\langle x \rangle \neq \bar{x}$.
If the system does not experience ``collective'' diffusion then $\bar{x}(t)=0$ and the modified MSD coincides with the conventional one:

$$\langle \Delta x^2 \rangle_{\text{corr}} = \langle x^2 \rangle.$$ 

The diffusion curves calculated by using the modified MSD are presented
in Fig.~\ref{FigSFDSC}(b).
For $P=0$, the diffusion curve (shown by black open squares) after the subdiffusive
regime reaches saturation (i.e., $\langle \Delta x^{2}(t) \rangle$$_{\text{corr}}$ = const).
The observed behavior is similar to that of a finite linear chain with fixed ends
(see Ref.~\cite{PRE80-051103}).
For $P\neq0$, all the diffusion curves in the long-time limit are characterized
by $\alpha = 1.0$, \textit{independent} of the value of the probability $P$,
as seen in Fig.~\ref{FigSFDSC}(b).
In other words, the long-time diffusion does {\it not} depends on the probability
of mutual exchanges of particles and has the same long-time behavior for {\it any}
probability $P\neq0$.
Here we would like to emphasize again that the long-time behavior of the diffusion curves
is free from the ``collective'' diffusion effect and is only determined by particle
jump diffusion.
Increasing a number of sites in the model corresponds, in fact, approaching to the model of infinite chain. We have found that the increasing a number of sites leads to growth of the $\langle \Delta x^{2}(t) \rangle$$_{\text{corr}}$ limit of saturation, on the one hand, and to a shift of $t_{\text{chain}}$ to larger $t$, on the other hand. Hence, extrapolating our results to the case of infinite chain, we can conclude that in this case as well as in the case of finite-size chain, the breakdown of single-file condition leads to an abrupt transition from subdiffusive to the normal diffusion regime.

The difference in the diffusive curves is just the time $\tau_{\text{tran}}$ from subdiffusive regime to the normal regime: for low $P$ it ($\tau_{\text{tran}}$) is long enough while for high $P$ it ($\tau_{\text{tran}}$) is short.
It is easy to see that $\tau_{\text{tran}} \sim 1/P(\%)$. 
Thus, we can conclude that in the long-time limit the transition from $~t^{0.5}$ to $~t^{1.0}$ behavior is {\it abrupt}. 
Note that our calculations performed using the modified MSD $\langle \Delta x^{2}(t) \rangle$$_{\text{corr}}$  reproduce the results of Ref.~\cite{PRE80-051103} for a closed ``box''. 
This is explained by the fact that in the closed ``box'' geometry the
center of mass (or collective) diffusion is zero, and it is natural that the roughness (see Ref.~\cite{PMCentres}) and the particles diffusion coincide.

%********************************************************************
\section{Conclusions}

We have studied a monodisperse system of interacting particles
subject to three types of confinement potentials:
(i) a 1D hardwall potential, (ii) a 1D parabolic confinement potential which
both characterize a \textit{quasi}-1D system, and (iii) a circular confining
potential, which models a finite size system.
In order to study the diffusive properties of the system, we have calculated
the mean-squared displacement (MSD) numerically through molecular dynamics
(MD) simulations.
For the case where particles diffuse in a straight line in a
Q1D channel, different diffusion regimes were found for different values
of the parameters of the confining potential ($\chi$ or $R_{w}$).
%
%Initially, for times $t < t_{a_0}$ (normal diffusion regime (I)), particles diffuse freely, and therefore, the MSD $\langle \Delta x^{2}(t) \rangle$ scales with $t^{1.0}$. The characteristic time scale $t_{a_0}$ is the time needed for a particle to diffuse over a certain distance comparable to the inter-particle distance, $a_{0}$. For times $t_{a_0} < t < t_s$, there is a transient regime (II), which has a non-trivial functional form, where $t_s$ is the characteristic time scale when one particle starts to feel the potential created by the neighboring particles. For time scales $t_s < t < t_c$, we can see that there is an intermediate subdiffusive regime (III), where MSD $\langle \Delta x^{2}(t) \rangle$ scales with $t^{\alpha}$ (with $0.5 \leq \alpha < 1.0$) and $t_{c}$ is the characteristic time scale when the system starts to diffuse as a whole.
%
We have found that the normal
diffusion is suppressed if the channel width $R_{w}$ is between $0.20$
and $0.50$ (or by $2.0 < \chi < 3.5$, for the case of parabolic 1D confinement),
leading the system to a SFD regime for intermediate time scales.
For values of $R_{w} \geqslant 0.56$, particles will be able to cross each other
and the SFD regime will be no longer present.
%For the Q1D channel confined by a hardwall potential, a reentrant behavior is observed in the slope of the MSD curves for the case the mean-squared displacement is calculated for both directions ($x$ and $y$). Which means that the interparticle potential plays an important role in the diffusion properties of systems in confined geometries.

The case of a circular channel corresponds to, e.g., the set-up
used in experiments with sub-millimetric metallic massive balls
diffusing in a ring with a parabolic potential profile created
by an external electric field.
The strength of the potential (which determines the effective
``width'' of the channel) can be tuned by the field strength.
Contrary to the case of hard-wall confinement,
where the transition (regarding the calculation of the scaling exponent ($\alpha$) of the MSD $\langle \Delta x^{2}(t) \rangle$$\propto t^{\alpha}$) is sharp,
a smooth crossover between the 1D single-file and the 2D diffusive
regimes was observed.
This behavior is explained by different profiles for the
distribution of the particle density for the hard-wall and
parabolic confinement profiles.
In the former case, the particle density reaches its maximum
near the boundaries of the channel resulting in a massive
breakdown of the SF condition and thus in a sharp transition
between the different diffusive regimes.
In the latter case, on the contrary, the density distribution
function has a maximum near the center which broadens with
decreasing strength of the confinement.
This results in a smooth crossover between the two diffusion
regimes, i.e., SFD and 2D regime.
The analysis of the crossing events, i.e., the rate of the crossing events $\omega_{c}$ as a function of the confinement parameter $\chi$ or $R_w$, supports these results: the function $\omega_{c}(\chi/R_w)$ displays a clear signature of either ``smooth'' or ``abrupt'' behavior. 

We also addressed the case of a finite discrete chain of diffusing particles.
It was shown that in this case the breakdown of the single-file condition
(i.e., when the probability $P$ of particles bypassing each other is non-zero)
leads to an abrupt transition from a subdiffusive regime to the
normal diffusion regime.

\section*{Acknowledgments}

This work was supported by CNPq, FUNCAP (Pronex grant),
the ``Odysseus'' program of the Flemish Government, the
Flemish Science Foundation (FWO-Vl), the bilateral
program between Flanders and Brazil, and the collaborative
program CNPq - FWO-Vl.


\begin{references}
\bibitem{underdamped}
For underdamped systems, the initial fast growth of the MSD (i.e., ballistic regime) is characterized by $\alpha = 2$.
\bibitem{TAEA}
As recently shown when studying single-trajectory 
averages~\cite{pre85JeonMetzler}, time average (TA) may differ from ensemble average (EA) MSD not only for nonergodic processes (for example, for anomalous diffusion described by continuous time random walks (CTRWs) models~\cite{EBarkaiPRL,Klafter}), 
but also for some ergodic processes in small complex systems.
\bibitem{powerlaw}
This apparent power-law behavior is characterized by MSD $\langle \Delta x^{2}(t) \rangle$$\propto t^{\alpha}$, and it is an intermediate phenomena due to the interplay between the crossover from the MSD $\langle \Delta x^{2}(t) \rangle$$\propto t^{0.5}$ regime to $\langle \Delta x^{2}(t) \rangle$$\propto t^{1.0}$ regime. For details, see Ref.~\cite{RKutner}.
\bibitem{nonzigzag}
Note that this is not a zig-zag transition: the function $P_{\text{rad}}(r)$
still has a single maximum which is shifted from the center of the channel.
\end{references}
\end{document}